\documentclass[11pt]{article}
\pdfoutput=1

\usepackage{latexsym}
\usepackage{enumerate}
\usepackage{url}
\usepackage{float}
\usepackage[hypertexnames=false,hyperfootnotes=false]{hyperref}
\usepackage{texnansi}
\usepackage{color}
\usepackage{tikz}
\usepackage[margin=10pt,font=small,labelfont=bf]{caption}
\usepackage{graphicx}
\usepackage{changepage}
\usepackage[subrefformat=parens,labelformat=parens]{subcaption}
\usepackage{afterpage}
\usepackage{enumitem}
\usepackage[boxed]{algorithm}
\usepackage{algpseudocode}
\usepackage[normalem]{ulem}
\usepackage{lmodern}
\usepackage{booktabs}
\usepackage{sectsty}
\usepackage{ifthen}
\usepackage{amsmath}
\usepackage{amsthm}
\usepackage{amssymb}
\usepackage{amsfonts}
\usepackage{thmtools}
\usepackage{thm-restate}
\usepackage{mathtools}
\usepackage{xspace}
\usepackage{titling}
\usepackage[numbers]{natbib}
\usepackage{xfrac}
\usepackage{multirow}
\usepackage{bigdelim}
\usepackage{bm}
\usepackage{cleveref}
\usepackage{multirow}
\usepackage[textsize=tiny]{todonotes}
\declaretheoremstyle[headfont=\sffamily\bfseries,bodyfont=\itshape]{thm-sf}

\crefname{assumption}{assumption}{assumptions}

\renewcommand{\thmcontinues}[1]{\hyperref[#1]{continued}}
\newcommand{\paraheader}[1]{\smallskip\noindent{\sffamily\bfseries #1}}
\usetikzlibrary{arrows,patterns,plotmarks,pgfplots.groupplots}
\tikzstyle{every picture} += [>=stealth]
\tikzset{axis/.style={semithick, line join=miter}}
\allsectionsfont{\sffamily}
\makeatletter
\def\@seccntformat#1{\csname the#1\endcsname.\quad}
\makeatother

\floatname{algorithm}{\normalfont\sffamily\bfseries Algorithm}
\floatstyle{ruled}
\provideboolean{fastcompile}
\newcommand{\hidefastcompile}[1]{\ifthenelse{\boolean{fastcompile}}{}{#1}}
\usepackage{pgfplots}
\usepackage{pgfplotstable}
\usetikzlibrary{calc}
\usepackage{mathtools}
\definecolor{orange}{rgb}{0.85,0.33,0.13} 
\definecolor{green}{rgb}{0.13,0.85,0.33}
\definecolor{purple}{rgb}{0.33,0.13,0.85}
\definecolor{lime}{rgb}{0.65,0.85,0.13}
\definecolor{blue}{rgb}{0.13,0.65,0.85}
\pgfplotscreateplotcyclelist{tricolor}{%
  orange,every mark/.append style={fill=orange!80!black},mark=*\\%
  green,every mark/.append style={fill=green!80!black},mark=square*\\%
  purple,every mark/.append style={fill=purple!80!black},mark=otimes*\\%
  black,mark=star\\%
  orange,every mark/.append style={fill=orange!80!black},mark=diamond*\\%
  green,densely dashed,every mark/.append style={solid,fill=green!80!black},mark=*\\%
  purple,densely dashed,every mark/.append style={solid,fill=purple!80!black},mark=square*\\%
  black,densely dashed,every mark/.append style={solid,fill=gray},mark=otimes*\\%
  orange,densely dashed,mark=star,every mark/.append style=solid\\%
  green,densely dashed,every mark/.append style={solid,fill=green!80!black},mark=diamond*\\%
}
\pgfplotsset{colormap={tricolormap}{color=(orange) color=(green) color=(purple)},
  colormap={quadcolormap}{color=(orange) color=(lime) color=(blue) color=(purple)}}
\pgfplotstableset{%
  font=\small,
  every head row/.style={before row=\toprule[1pt], after row=\midrule},
  every last row/.style={after row=\bottomrule[1pt]}}
\pgfplotsset{compat=1.15}

\usepackage{setspace}
\usepackage[margin=1in]{geometry}

\provideboolean{submissionversion}
\setboolean{submissionversion}{true}
\provideboolean{blindversion}

\ifthenelse{\boolean{submissionversion}}{
  \renewcommand{\todo}[2][1]{}
  
  \newcommand{\deledit}[1]{}
}{
  
  \newcommand{\deledit}[1]{{\color{orange} \sout{#1}}}
}

\newcommand{\WETH}{\texttt{WETH}}

\title{Quantifying Price Improvement in Order Flow Auctions\thanks{We thank Dan Robinson at Paradigm for helpful discussions.}}
\author{Brad Bachu\thanks{Uniswap Labs, \texttt{brad.bachu@uniswap.org}.} \and Xin
  Wan\thanks{Uniswap Labs, \texttt{xin@uniswap.org}.} \and Ciamac C.~Moallemi\thanks{Columbia
    University, \texttt{ciamac@gsb.columbia.edu}. Moallemi is supported by the Briger Family
    Digital Finance Lab at Columbia Business School, and is also a research advisor to Paradigm
    and to other fintech companies.}}
\date{Initial verion: April 30, 2024 \\
  This version: May 22, 2024 }

\begin{document}
\maketitle

\begin{abstract}
This work introduces a framework for evaluating onchain order flow auctions (OFAs), emphasizing the metric of price improvement. Utilizing a set of open-source tools, our methodology systematically attributes price improvements to specific modifiable inputs of the system such as routing efficiency, gas optimization, and priority fee settings. When applied to leading Ethereum-based trading interfaces such as 1Inch and Uniswap, the results reveal that auction-enhanced interfaces can provide statistically significant improvements in trading outcomes, averaging 4-5 basis points in our sample. We further identify the sources of such price improvements to be added liquidity for large swaps. This research lays a foundation for future innovations in blockchain based trading platforms.

\end{abstract}

\onehalfspacing

\section{Introduction}

In recent years, blockchain-based Automated Market Makers (AMMs) have gained significant traction. Uniswap, the largest AMM by trading volume, has recently surpassed \$2 trillion USD in transactions \cite{uniswap2tvolume}. These systems, designed to automate asset trading, promise more efficient markets and potentially lower trading costs.

Despite their success, AMMs encounter notable challenges such as fragmented liquidity, susceptibility to adversarial strategies, and unpredictable trade executions. Studies estimate that inefficiencies and adversarial strategies have extracted over \$540 million USD from the AMM ecosystem over 32 months  \cite{qin2022quantifying}. 

To address these issues, Order Flow Auctions (OFAs) have been proposed as a solution. Protocols such as 1inch Fusion \cite{1inchfusion}, UniswapX  \cite{uniswapx},  CowSwap \cite{cowswap} and MEV Share  \cite{mevshare} introduce methods like batching, auctioning, and matching of orders to optimize trading outcomes for all participants. Specific implementations of OFAs differ \cite{frontierofasummary}, but in general they shift the complexities of optimizing swap outcome to specialized participants, and redistribute the surplus back to relatively unsophisticated swappers. Proponents of OFAs claim that they could provide price improvements, as well as more secure and predictable trading environments \cite{ofafb}. However, empirical validation of these benefits has remained absent due to the complexity of integrating blockchain technology with financial market dynamics, until now.

This paper introduces a new, open-source methodology to evaluate and compare OFAs, focusing on price improvements (PI). We decompose price improvement into three controllable factors by the OFA provider: routing efficiency, gas optimization, and priority fee settings. This decomposition allows for detailed analysis and potential optimization insights for OFA systems.

We apply this methodology to analyze the top two OFA providers by volume and trade count — 1Inch and Uniswap — on the Ethereum mainnet \cite{ofainterfaces}, which hosts the majority of onchain trading activities. In addition, both of these interfaces utilize similar mechanisms to improve prices, and so are a good starting point to test the methodology and interpret the results\footnote{Other interfaces that utilize batch auctions and rebate mechanisms require special treatment for a consistent baseline, and so are left for future work.}. Preliminary results suggest that OFAs, particularly those utilizing RFQ (request-for-quote)-informed onchain Dutch auction systems, improve trading execution. Additionally, our attribution model provides insights into when one OFA outperforms another. This work lays the foundation for future analysis such as batch auctions and rebate systems.

\paragraph{Summary of our contributions.}
In this paper, we make four main contributions:
\begin{itemize}
    \item \emph{Price Improvement Definition And Computational Framework.} 
    We introduce a framework to evaluate OFA execution quality, focusing on price improvement — the difference between the realized price of a swap and its counterfactual price. We formally define price, price improvements, and counterfactual price in a way that is generalizable across different OFA implementations. 
    We also provide a way to consistently generate counterfactual prices for any realized swap using fully open-source code and data source.  

    \item \emph{Methodology For Gas Cost Internalization.} 
    Research indicates that gas costs can constitute a substantial portion of the effective spread in AMM trades, with \cite{adams2023costs} noting it could exceed 90\% for small-sized ones. However, current literature lacks a formalized approach to factor gas costs into execution quality assessments with statistical precision. Our framework introduces a method for integrating gas costs into trade prices (``\textit{gas internalization}''). Our method has two advantages compared to common methods used by practitioners. Compared to using the average or median gas cost of similar transactions as the benchmark, our method exposes more variability and controls for more granular trade-offs. In addition, we show that gas cost in simulated transactions, even in the same block, will have a statistical bias, which our method corrects for.

    \item \emph{Price Improvement Attribution Model.}
    We propose a model that attributes price improvements to various controllable inputs using a Taylor Series expansion. Our empirical analysis shows that non-attributable differences are minimal for the interfaces studied. Our attribution model provides OFA implementers with direct insights into how they could improve their systems.

    \item \emph{Empirical Application.} 
    We empirically test our methodology on historical data from leading OFA interfaces on the Ethereum blockchain. The findings confirm our method's capability to discern execution quality across systems with statistical precision. Results show that dutch auction-based OFAs yield an average price improvement of 4 to 5 basis points for users, with this improvement being significant across various trade sizes—though it's noted that smaller trade sizes exhibit variability that may not be statistically distinct from zero. The attribution model points to increased liquidity and optimized routing as the primary contributors to price improvement in larger trades. Conversely, for smaller trades, efficient gas usage and selection of pools with lower gas costs emerge as the key factors.

\end{itemize}

Our work provides, to the best of our knowledge, the first formal definition and framework for assessing price improvements in OFAs, offering granular insights into their operational effectiveness. Preliminary empirical findings suggest that OFAs could outperform existing interfaces that solely rely on onchain data and liquidity sources.


\section{Literature Review}\label{sec:literature}

Execution quality in financial markets has been a significant topic in previous literature. \cite{bessembinder2010bid} reviews methods for measuring execution cost, such as quoted spread, effective spread, and realized spread. These metrics, although straightforward, may vary depending on measurement techniques. For instance, \cite{bessembinder2003issues} demonstrates how the methodology for comparing trades to publicly-available quotes can influence the reported execution costs.

Several empirical studies have quantified execution costs within the US equities market. \cite{wah2017comparison} evaluates execution quality across US exchanges, while \cite{bessembinder1997comparison} compares average execution costs for large, medium, and small capitalization stocks on the New York and NASDAQ stock markets during 1994.

Policy changes and market structure modifications are often followed by detailed analyses aimed at evaluating their impact on market efficiency and execution costs. \cite{bessembinder2003trade} examines trade execution costs and market quality for NYSE and NASDAQ stocks before and after the switch to decimal pricing in 2001, finding a reduction in execution costs. Similarly, \cite{bessembinder1999trade} investigates how execution quality varied between the NYSE and NASDAQ following SEC changes to order handling rules and tick size reductions.

Other factors also play a role in determining execution costs. \cite{sauglam2019short} finds that short-term trading skills of investors can influence the execution quality they experience.

Moreover, execution quality encompasses more than just cost. \cite{boehmer2005dimensions} argues that it is multi-dimensional, with speed being an important factor alongside cost. \cite{engle2006measuring} models the trade-off between lower execution costs and faster execution speeds. \cite{garvey2009intraday} studies intraday execution quality patterns on Nasdaq stocks, using proprietary order-level data to identify compensatory patterns between speed and cost.

The brokerage setups and order routing decisions also affect execution. The concept of price improvement, which is the difference between the executed and quoted prices, is crucial in assessing the effects of factors like payment for order flow or differential pricing. \cite{brolley2020price} investigates price improvement and execution risks in lit and dark markets. \cite{angel1994gets} examines instances when NYSE market orders are executed at prices better than the current specialist quotes. \cite{rhodes2005price} introduces a theory where price improvements result from negotiations and vary with customers’ market power. \cite{desgranges2005reputation} explores how dealer reputation influences price improvement and its subsequent effects on market liquidity and traders’ welfare. \cite{parlour2003payment} develops a model of price competition in broker and dealer markets, incorporating payment for order flow. Lastly, \cite{ernst2022payment} documents how payment for order flow, spreads, and price improvements vary across different asset classes.

Execution cost has also been studied in the context of cryptocurrency markets, including onchain markets. \cite{adams2023costs} examines the trading costs on the Uniswap Protocol, reporting that the effective spreads for the top trading pair is similar to that of traditional asset classes. \cite{wan2022just} investigates strategies of just-in-time liquidity provision and their effects on execution quality for individual trades. Additionally, \cite{capponi2023paradox} develops a model to analyze the influence of just-in-time liquidity on both long-term displayed liquidity and overall execution costs. Finally, \cite{chitra2024analysis} discusses potential trade-offs in OFA designs. 

To our knowledge, there has been no development of a theoretical framework or conduct of empirical studies that compare execution quality across various onchain trading platforms and interfaces. Such comparisons would need to be economically consistent across different order flow auction implementations and include all factors affecting the net economic outcomes of trading, particularly considering transaction gas costs. Our framework and empirical analysis in this paper aim to bridge this gap.

\section{Theoretical Framework} \label{sec:theory}

In our research, we aim to define `price' and `price improvement' consistently across different OFA systems. Transaction cost (gas fee) handling is the main point of difference across various OFAs that introduces complexities into our comparison. For example, some OFAs enable users to place orders without paying gas fees upfront. The gas fees are then paid by another party, known as the filler or solver, who takes part of either the input token or the output token as compensation for gas fee. Other OFAs require their users to pay gas fees themselves. These differences make it challenging to compare prices between OFAs, since gas fee could be denominated in a different unit than the input token (i.e., the users could be effectively paying two tokens for one trade), and exchange rate at which we should be converting gas fee into either the input or output token is not obvious. On the other hand, ignoring the gas cost would lead to incorrect results. As we will show later, sometimes the difference in gas cost is be enough to make one OFA better than the other, despite giving less output token in the trade before gas cost.

Our methodology is applicable to transactions where either the input or the output token is either the gas token, or a wrapped version of the gas token (in the case of Ethereum, ETH or WETH). These transactions currently account for about 90\% of all DEX trades on Ethereum mainnet (or ``Ethereum'').\footnote{https://dune.com/queries/3675220} This helps us standardize how we internalize gas fees. By doing this, we can apply our price definitions uniformly to all OFAs, whether they cover gas costs for their users or not. In the next section, we formally define ``price'' that internalizes gas cost of the transaction.

\subsection{Price}

Consider a scenario where a user has a pre-trade balance of two tokens, A and B, denoted as $(a,b)$, and a post-trade balance of $(a',b')$, after trading token A in exchange for token B. The price $p$ is defined by
\begin{equation} \label{eq:price_def}
    p = \frac{b' - b}{a - a'},
\end{equation}
where the signs are chosen to reflect that the amount of Token B is increasing and the amount of Token A is decreasing. As mentioned, since we restrict one of A or B to be either WETH/ETH for transactions on Ethereum mainnet, our definition takes into account the transaction fee of the trade, which is usually paid in ETH.

In the context of onchain trading, the balance changes described in \eqref{eq:price_def} are a result of providing some input token $i$ and gas $g$, and getting some output token $o$. For example, consider a scenaro where a user trades $i$ amount of ETH, paying gas $g$ ETH, for $o$ amount of USDC.
The user's balance in USDC increased by $o$, and that of ETH decreased by $i + g (b+f)$, where $b$ and $f$ represent the base fee and priority fee, respectively. This is reflected in \eqref{eq:price_def} as,
\begin{equation}
        p =  \frac{b' - b}{a - a'} = \frac{o}{i + g(b + f)} \,.
\end{equation}
Other scenarios are described in detailed in \Cref{sec:price_details}.

The price in our context is influenced by several factors. For OFAs where gas is internalized, such as 1Inch-Fusion and UniswapX, the factors include simply the amount of output tokens $o$ and the amount of input tokens $i$. In cases where gas is not internalized, such as 1Inch Aggregator and Uniswap Classic, additional factors come into play, including the gas used $g$, the base fee per gas $b$, and the priority fee per gas $f$.

It is important to note, however, that in scenarios where a user initiates a transaction with a fixed quantity of input tokens $i$, aiming for settlement within a given block with a predetermined base fee per gas $b$, the variables that an OFA can manipulate are limited to the output token amount $o$ (through routing through different liquidity pools), the gas used $g$ (through doing different sets of operations in the transaction), and the priority fee per gas $f$ (through changing the transaction setting). To facilitate our analysis, we introduce a vector $\vec{x} = (o, g, f)$ that aggregates these variables. This allows us to succinctly define the price function as
\begin{equation}
    p := p(o, g, f) = p(\vec{x})\,.
\end{equation}

\subsubsection{Example - Price In Transactions With Gas Cost}
Traditional transactions, such as those executed through Uniswap Classic or the classic 1Inch Aggregator interface, typically involve users paying gas in ETH. When ETH happens to also be the input token in the swap, the user's ETH balance decreases by the sum of the input token amount plus the gas cost, while the output token balance increases by the output token amount. The price in these instances is calculated as the ratio of the output token amount to the sum of the input token amount and gas cost. Conversely, if ETH is the output token, the ETH balance increases by the net amount of the output tokens after subtracting the gas cost. The input token balance decreases by the amount of the input tokens used. The price is then determined as the ratio of these two amounts. In scenarios where the output in ETH is less than the gas cost, the price could appear negative, indicating a net loss in both token balances.

\subsubsection{Example - Price In ``Gas-free'' Transactions}
A newer type of transactions facilitated by Permit2 signatures\footnote{https://blog.uniswap.org/permit2-and-universal-router} involves users authorizing others to move their tokens under certain conditions without upfront gas payment. A filler then executes the transaction, covering the gas costs and transferring the output tokens to the user, subsequently claiming the input tokens as compensation. In this scenario, the user's wallet shows only the net increase in output tokens and decrease in input tokens without any change in gas token balances. Thus, the price is straightforwardly calculated as the negated ratio of the increase in output tokens to the decrease in input tokens.

\subsubsection{Example - Price In Transactions With Potential MEV Rebate}
Transactions involving potential MEV rebates are more complex, often encompassing several linked transactions. Platforms like MEV Share and MEV Blocker facilitate this process, where a potential rebate is provided back to the user via an ETH transfer. Despite the initial gas costs paid by the user, the effective gas cost may be offset by the MEV rebate, potentially resulting in a negative net gas cost. The pricing mechanism in these cases resembles the traditional transactions but adjusted for the possible reduction in gas costs due to the MEV rebate.

\subsection{Price Improvement}
We define the price improvement $\pi$, as the relative difference between the realized price $p$, and some baseline price $p'$, as
\begin{equation}
        \pi(p,p^\prime) = \frac{p - p^\prime}{p^\prime} \,.\label{eq:pi_1}
\end{equation}
We define the baseline price, $p'$, through a baseline vector $\vec{x}' = (o', g', f')$, encapsulating alternative states of the variables under consideration. This leads to the definition:
\begin{equation}
    p' = p(o', g', f') = p(\vec{x}'),
\end{equation}
The baseline prices should be interpreted as a counterfactual price that could be reasonably achieved by a typical user if they did not use the OFA. We will primarily use calls to the Uniswap Classic routing API with historical state data to generate $\vec{x}'$ as a counterfactual baseline. How we generate a consistent $\vec{x}'$ is the focus of \Cref{sec:method}.

\subsection{Price Improvements Across Time}\label{subsec:PriceImprovementsAcrossTime}
Our primary definition of price improvement $\pi$ involves comparisons with simulations against the settlement time $t_0$. However, we will also find it useful to evaluate price improvement across various time offsets $\Delta t$ from $t_0$. In doing so, we can evaluate the impact of execution speed and account for any timing decisions / differences among OFAs.

As mentioned, implicit in the definition of realized price $p$ is the settlement time $t_0$, $p = p(t_0)$. However, a counterfactual price $p'$ can be generated at any time $t$, $p' = p'(t)$.
In this way, we can include the time dimension in our definition price improvement as
\begin{align}
    \pi(p,t_0; p',t) &= \frac{p(t_0) - p'(t)}{p'(t)}\,.
\end{align}
Since the settlement time $t_0$ is fixed historically, the only meaningful differences are offsets from the settlement time $\Delta t = t -t_0$. To explore this, can apply a shift of $-t_0$ to our definition, so that we are always considering times relative to settlement time, effectively, setting $t_0 =0$. Thus, $\pi(p,t_0; p',t)\rightarrow\rho(p,0; p',t-t_0) = \rho(p,0;p',\Delta t)$.
This yield the more practical definition,
\begin{align}
    \rho(p;p',\Delta t) &=\frac{p - p'( \Delta t)}{p'(\Delta t)}  \,,
\end{align}
where we have dropped the $0$ for simplicity.
To consider the price improvement at the settlement time, simply set $\Delta t = 0$, yielding our original definition \eqref{eq:pi_1},
\begin{align}
    \rho(p;p',0)  = \pi(p,t_0; p',t_0)\,. \label{eq:pi-mo}
\end{align}




There are several reasons why this curve $\rho(p;p',\Delta t)$ is informative:
\begin{enumerate}
    \item It helps account for potential differences in the speed of transaction inclusion between the counterfactual interface and the actual interface used, highlighting any robustness issues related to execution speed.
    \item Many trading interfaces incorporate mechanisms that affect how or whether orders are filled. By examining multiple offsets, we can identify and adjust for any selection biases these mechanisms might introduce.
    \item Blockchain environments can be unpredictable, with factors such as unexpected gas spikes or specific contract storage issues affecting outcomes. By assessing a range of blocks, we mitigate the impact of such variables, ensuring our findings are not skewed by transient blockchain conditions.
    \item Transaction ordering, especially for Ethereum mainnet, is highly adversarial. It is often in the interest of the block builders to place a transaction in a position that is against the interest of the transaction sender. By comparing against multiple probable counterfactuals in terms of ordering, the curve would highlight any potential sensitivity against transaction ordering.
    \item The time offset to calculate $\rho$ varies depending on the desired interpretation, for example, comparing counterfactual price at arrival or counterfactual price at execution. By comparing to the most relevant times, we can facilitate all interpretations.
\end{enumerate}

\subsection{PI Attribution}\label{subsec:attribution}

In our analytical framework, PI is conceptualized as an aggregated result of various underlying decisions, such as routing, gas usage, priority fee settings, and many others. However, it lacks interpretability. To provide a more granular insight into how PI is achieved across different systems, we break down PI into three components, each with economic significance. We describe $\pi$ as a sum of OFA controllable (directly or indirectly) quantities: routing optimization $\pi^{\text{routing}}$, gas optimization $\pi^{\text{gas}}$ , and priority fee optimization $\pi^{\text{fee}} $:

\begin{equation}
        \pi = \pi^{\text{routing}} + \pi^{\text{gas}} +\pi^{\text{fee}} \,, \label{eq:pi_attribution}
\end{equation}where $\pi^{\text{routing}}$ represents the price improvement through optimizing liquidity access, $\pi^{\text{gas}}$ captures the savings from reduced gas costs, and $\pi^{\text{fee}}$ accounts for the impact of different priority fees.

Interface decision-making impacts these PI components through several mechanisms:
\begin{enumerate}
    \item \textbf{\sffamily Route Optimization}:
    The routing decision by an interface involves selecting which liquidity sources to use, which may include on-chain liquidity pools or off-chain sources. The selection process must balance the benefits of additional liquidity against the increased gas costs associated with incorporating more liquidity sources. In theory, more liquidity would always be preferable if it did not incur additional gas costs. However, in practice, each additional liquidity source increases the transaction's gas cost, which may negate the benefits of the added liquidity. The component $\pi^{\text{routing}}$ captures the amount of PI achieved through including more liquidity in the route than the baseline.

    \item \textbf{\sffamily Gas Efficiency}:
    Optimizing gas efficiency is closely tied to the routing decision. An effective routing algorithm not only selects the optimal liquidity sources but also minimizes the transaction's gas usage. This involves a trade-off between accessing sufficient liquidity and managing the transaction's cost-effectiveness. The challenge is to achieve the best possible output for the input while keeping gas expenses as low as feasible. The component $\pi^{\text{gas}}$ captures the amount of PI achieved through using less gas than the baseline.

    \item \textbf{\sffamily Priority Fee Setting}:
    The setting of priority fees involves recommending a default fee that influences the speed and likelihood of a transaction's inclusion in the blockchain. Under the current gas fee mechanism (EIP-1559), any positive priority fee is typically sufficient for inclusion. Despite this, many interfaces recommend a priority fee that is significantly above zero to achieve faster execution speeds. This decision can impact the price improvement negatively, as higher fees might offset the gains made through efficient trading and routing strategies. The component $\pi^{\text{fee}}$ captures the amount of PI achieved through paying less priority fee than the baseline.

\end{enumerate}

To do the attribution, we approximate \eqref{eq:pi_attribution}.
We begin by Taylor expanding $p(\vec{x})$ about the baseline variables $\vec{x}'$,
\begin{align}
    p(\vec{x}) &= p(\vec{x}') + \left.\frac{\partial p}{\partial o}\right|_{\vec{x}'} (o - o') + \left.\frac{\partial p}{\partial g}\right|_{\vec{x}'} (g - g') + \left.\frac{\partial p}{\partial f}\right|_{\vec{x}'} (f - f') + R(\vec{x},\vec{x}')\,,
\end{align}
where $R(\vec{x},\vec{x}')$ represents the remainder term in the Taylor expansion.
The quantity of interest to us is price improvement. Rearranging the above, we have
\begin{alignat}{4}
        \pi &= \left.\frac{\partial p}{\partial o}\right|_{\vec{x}'} \frac{(o - o')}{p'} &&+ \left.\frac{\partial p}{\partial g}\right|_{\vec{x}'} \frac{(g - g')}{p'} &&+ \left.\frac{\partial p}{\partial f}\right|_{\vec{x}'} \frac{(f - f')}{p'} &&+ \frac{R(\vec{x},\vec{x}')}{p'} \\
            &= \pi^{\text{routing}}_0 &&+ \pi^{\text{gas}}_0 &&+ \pi^{\text{fee}}_0 &&+ \pi^{\text{remainder}},
\end{alignat}
where the $0$ subscript is used to represent the leading order contribution to each corresponding term.
Of course, it is not guaranteed that remainder term $\pi^{\text{remainder}}$ is small, but we
will find that for some cases, it is.

Note that for OFAs where gas is internalized, the amount output token $o$ that the user receives is the amount after all transaction fees have been paid. Since onchain data provides $g$, $b$ and $f$, we can actually calculate the amount of token a user would have received pre-transaction fee. We can utilize this strategy to attribute $\pi$ for such OFAs.


\section{Methodology}\label{sec:method}
To demonstrate the efficacy of our proposed methodology, we apply it to the top two Order Flow Auction (OFA) interfaces based on market volume: the 1Inch Interface and the Uniswap Interface. We describe what data is collected, the process of generating baselines, calibrations to the generator and uncertainty analysis.

\subsection{Sample Interfaces}
In this section, we will describe the mechanisms behind these trading interfaces.

The Uniswap Interface splits incoming orders into two distinct paths: Uniswap Classic and UniswapX. Historically, Uniswap Classic was the only option. When users input both tokens and the specified amount (either input or output), the interface calls an open-source routing API to identify the best route across all Uniswap pools. The interface then displays the optimal swap, which the user can confirm by clicking the `Swap' button. Upon confirmation, the route information is converted into calldata, prompting the user's wallet to sign and broadcast the transaction to the blockchain network.

With the introduction of UniswapX in July 2023, users gained an alternative path upon visiting the interface. After entering both tokens and one of the amounts, the interface simultaneously queries the routing API and a set of specialized participants known as quoters. The quoters provide competitive quotes for the unspecified amount, and the interface determines whether the routing API's quote or a quoter's offer provides a better deal from the user's perspective. If the API's quote is superior, the flow defaults to Uniswap Classic; otherwise, it routes to UniswapX.

In the UniswapX flow, once users opt to proceed with the best quote from the quoters by clicking the `Swap' button, they sign an offchain message detailing the trade, the quote, and the quoter's identity. This message also includes contingencies for a Dutch auction should the order not fill promptly. The signed message is then disseminated through an API endpoint accessible to subscribed parties, with the initial order-filling advantage given to the best quoter.

Subsequently, any participant wishing to fill the order submits a transaction that adheres to the predefined conditions. A validation contract checks these conditions, and if met, the transaction proceeds successfully. For exact details of dutch orders, please refer to the whitepaper of UniswapX.\footnote{https://uniswap.org/whitepaper-uniswapx.pdf}

The 1inch Interface has a similar setup, where user orders are split between 1inch Aggregator (similar to Uniswap Classic) and 1inch Fusion (similar to UniswapX).

\subsection{Selection of Baseline}

Price improvement, as a measure, depends on the relevance and effectiveness of the baseline against which it is compared. In this section, we delineate the rationale behind our chosen baseline. Note that while our methodology utilizes this specific baseline, it remains flexible and can be adapted to alternative baselines by other researchers.

Our baseline for assessing any completed swap via a trading interface is the simulated outcome of that swap. This simulation involves two steps: firstly, obtaining a trade route by submitting a swap request with the same token pair and an exact-in amount equal to the completed swap's token in amount through the Uniswap open-source routing API; secondly, simulating the execution of this route at the end of the same block in which the actual swap was settled (referred to as the `settlement block') using the Tenderly Simulator. Use of this simulator is extremely important for accurate gas usage comparisons.

We pick this baseline for several reasons. Primarily, the Uniswap routing API aggregates liquidity from all Uniswap pools, which constitute a significant portion of total onchain liquidity.
Most aggregator trades optimizing for output are routed through these pools.
Furthermore, the API underpins the Uniswap Interface — a major platform by user count and transaction volume — mirroring a typical user experience that is accessible to the average swapper. Additionally, since the API is open to the public, any third party can replicate this analysis. We focus on end-of-block simulation both for easy of the simulation and for the fact that a counterfactual transaction could have actually been placed at the end of the block, while there is significant contention and competition for top-of-block transactions. Lastly, although using the mid-price from a centralized exchange could potentially offer a less blockchain-state-influenced benchmark, it fails to accommodate long-tail tokens, which, though rarely traded on centralized exchanges, often have pools on Uniswap. Moreover, utilizing a centralized exchange’s mid-price would preclude detailed analysis of onchain execution enhancements like routing efficiency, gas optimization, and priority fee settings. These are specific to onchain trading, and insights into them will be helpful for any onchain trading platform provider seeking to improve their execution quality.

Consequently, we selected the simulation against the settlement block via the Uniswap open-source routing API and Tenderly \cite{Tenderly2024} as our empirical baseline. A comprehensive explanation of the methodologies and inputs used in generating the simulation results is available in \Cref{sec:simulation_process}. 
 In the context of \Cref{subsec:PriceImprovementsAcrossTime}, we select the counterfactual price $p'(t_0)$ at the end of the execution block to perform the main analysis. However, to facilitate other interpretations, we consider alternative baselines at different time offsets (this also allows the consideration of a top-of-block baseline, since execution at the bottom-of-block at offset $-1$ is roughly equivalent to execution at the top-of-block at offset $0$).

\subsection{Data Collection}

In this section, we describe the procedure of collecting the historical values for input amount, output amount, gas used, base fee per gas, priroity fee per gas $(i,o,g,b,f)$ as well as the method of generating baseline values $\vec{x}' = (o',g',f')$.

We also restrict our data to WETH-USDC trades in November and December 2023.
All historical data is collected using Dune Analytics. Data from Uniswap Classic is collected by taking transactions that have used the Uniswap Classic routers, and filtered for Uniswap Interface users. Data from UniswapX is collected by taking transactions that have interacted with the UniswapX contract. Data for 1Inch is selected by using the \texttt{oneinch} table in Dune Analytics. All interface fees are ignored to focus on execution quality of the transactions.

\begin{table}[H]
\centering
\begin{tabular}{lp{2cm}p{2.2cm}p{2.2cm}p{2.2cm}p{2.2cm}}
\toprule
  \textbf{\sffamily Interface}
  & \textbf{\sffamily Settlement Path}
  & \textbf{\sffamily Sample Size (\# Swaps)}
  & \textbf{\sffamily \% of Parent Interface Swap Count}
  & \textbf{\sffamily Total Volume (\$)}
  & \textbf{\sffamily \% of Parent Interface Volume} \\ \midrule
\multirow{2}{*}{1Inch}              & Aggregator               & 1687                           & 36\%                                         & 37,891,096                 & 22\%                                    \\
           & Fusion                   & 2941                           & 64\%                                         & 134,221,271                & 78\%                                    \\ \midrule
\multirow{2}{*}{Uniswap}          & Classic                  & 1809                           & 16\%                                         & 28,573,360                 & 13\%                                    \\
            & X                        & 9607                           & 84\%                                         & 185,599,214                & 87\%                                    \\ \bottomrule
\end{tabular}
\caption{Distribution of swaps and volumes across different settlement paths for 1Inch and Uniswap interfaces.}
\label{tab:sample}
\end{table}

\Cref{tab:sample} shows how number of swaps and USD volumes are distributed across interfaces and their respective settlement paths. Note that both 1Inch and Uniswap have a relatively dominant settlement path - Fusion for 1Inch and UniswapX for Uniswap - and that the concentration is even more pronounced when we look at distribution by USD volume. This is consistent with prior expectations, since these OFA systems theoretically integrate more liquidity sources and providers, and could provide price improvement, especially for trades with large sizes.

Next, we set out to generate a baseline price $p'$, by generating baseline variables $\vec{x}' = (o',g',f')$. We do this via API calls for historic transactions through Uniswap Classic. For a given block time $t$, and input amount $i$, we can generate a counterfactual Uniswap Classic transaction that yields estimate of the output token $o'$ and gas used $g'$. We can summarize this by introducing a basline function $\mathcal{B}$ that maps historic $i,t$ to the counterfactual baseline $o',g'$,
\begin{equation}
        \mathcal{B}: (i,t) \rightarrow (o',g') \,,
\end{equation}
we set the baseline priority fee per gas $f'$ to be a consistent $0.1$ Gwei.

\subsection{Baseline Gas Corrections}
\begin{figure}[h]
    \centering
    \includegraphics[width=0.8\linewidth]{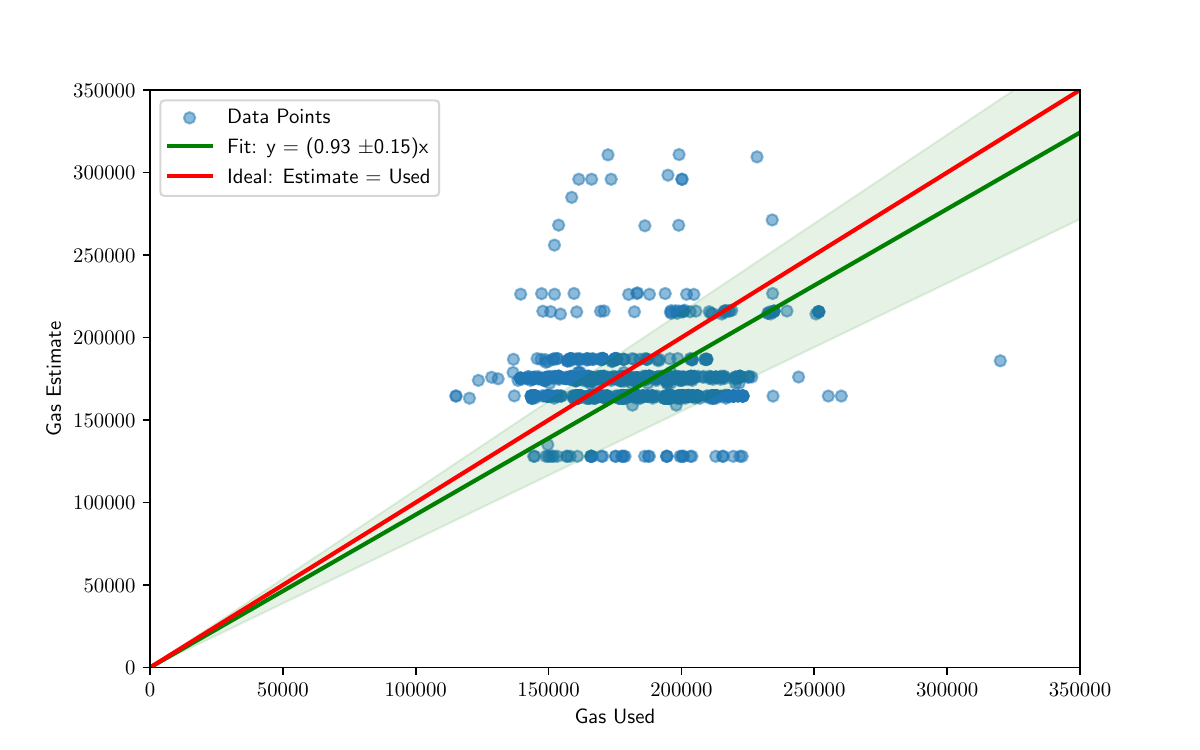}
    \caption{The gas use estimate $g'$ versus the actual gas used $g$ for Uniswap Classic transactions. The red line illustrates a perfect gas simulation, where the gas use estimate is the same as the gas used. However the discrepancy illustrates the need to calibrate the gas estimator. The green dashed line indicated the calibration, with the band indicating the level of uncertainty associated with the estimator.}
    \label{fig:gas_calibration}
\end{figure}

A major source of systematic error in our baseline is in the generation of gas estimations $g'$. To illustrate, \Cref{fig:gas_calibration} shows the variation between $g$ and $g'$ for Uniswap Classic transactions in our dataset. By definition, we would like this to be mostly $0$, since a counterfactual Uniswap Classic simulation of a historical Uniswap Classic transaction should be almost identical. However, in some cases, it is expected that the amount of gas you use is less than the amount of gas you are estimated to use, for example, through JIT \cite{wan2022just}. In addition, since the transaction is usually intra-block, and we can only simulate at the bottom of the block, we do not expect these numbers to align exactly. However, we will parametize our uncertainty in this estimation.

We correct for this systematic bias by regressing gas estimates for  historical Uniswap Classic transactions on the gas used, i.e., the regression specification $g' \approx \beta_1 g $
where $(\beta_1)$ is the estimated regression parameters.
Then, when we generate $g'$ for a counterfactual transaction, we can correct our bias by $g' \leftarrow g'/ \beta_1$.


\subsection{Uncertainty Analysis}\label{sec:uncertainty}

For price improvement $\pi$, and its attributes $\pi_0^{\text{attribute}}$, we compute an average $\bar{\pi}$ and $\bar{\pi}_0^{\text{attribute}}$
weighted by its estimated USD size. This value has two significant uncertainty contributions, a statistical $\sigma_{\text{stat}}$ and a systematic $\sigma_{\text{sys}}$ uncertainty. Thus, for any $x\in\{\pi, \pi_o^{\text{attribute}}\}$, we must compute $\bar{x} \pm \sqrt{ \sigma_{\text{stat}}^2 + \sigma_{\text{sys}}^2}$.

The statistical uncertainty is determined by computing the weighted standard error on the weighted mean. For a transaction $i$, and given $x_i \in \{\pi ,\pi_0^{\text{attribute}} \}_i$ with USD weight $w_i$,
the statistical uncertainty on the weighted mean $\bar{x}$ is determined as
\begin{align}
    \sigma_{\text{stat}}^2 &= \frac{\sum_i w_i (\bar{x} - x_i)^2}{n \sum_j w_j}\,. \label{eq:stat_uncertainty}
\end{align}

The systematic uncertainty on the gas estimate $\delta g'$, which derives entirely from the gas estimation, needs to be propagated from the uncertainty in the fit parameter $\delta\beta_1$ from \Cref{fig:gas_calibration}. We can compute these asymmetrically by varying the gas by $\delta\beta_1$. For example, $\sigma_{\text{sys}}^{\text{upper bound}} =\bar{x}(\beta_1 +\delta \beta_1) - \bar{x}(\beta_1 )$ and $\sigma_{\text{sys}}^{\text{lower bound}} =\bar{x}(\beta_1 ) - \bar{x}(\beta_1 - \delta \beta_1 )$.

\subsection{Uncertainty and Limitations}

The major source of uncertainty in this analysis comes from determining an accurate gas estimation. The systematic uncertainty comes entirely from the gas estimation. Ideally, in \Cref{fig:gas_calibration}, all points should lie along the line gas estimate = gas used. This is not the case due to a variety of reasons, most notably, the algorithm behind the simulation process described in \Cref{sec:simulation_process} has changed. Secondly, intra-block dynamics are not captured since we can only currently simulate at the end a block. Lastly, there is the added feature that, if your transaction used a Uniswap pool, and you simulate your transaction against the same pool, you actual transaction has affected (front-run) your simulation.

We plan on overcoming these challenges by investigating the potential to have a consistent routing algorithm, intra-block simulations, and exporing the possibility of removing transactions from the block before simulating.


\section{Results}\label{sec:results}

\begin{figure}[ht]
    \centering
    \begin{subfigure}{0.49\textwidth}
    \centering
        \includegraphics[width=\textwidth]{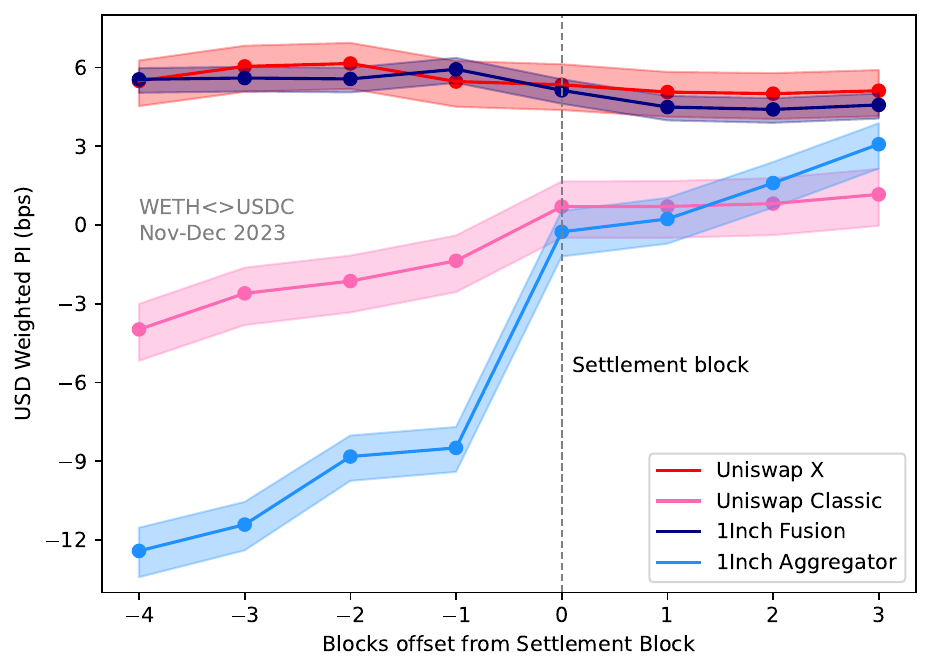}
    \end{subfigure}
    \begin{subfigure}{0.48\textwidth}
        \centering
        \includegraphics[width=\textwidth]{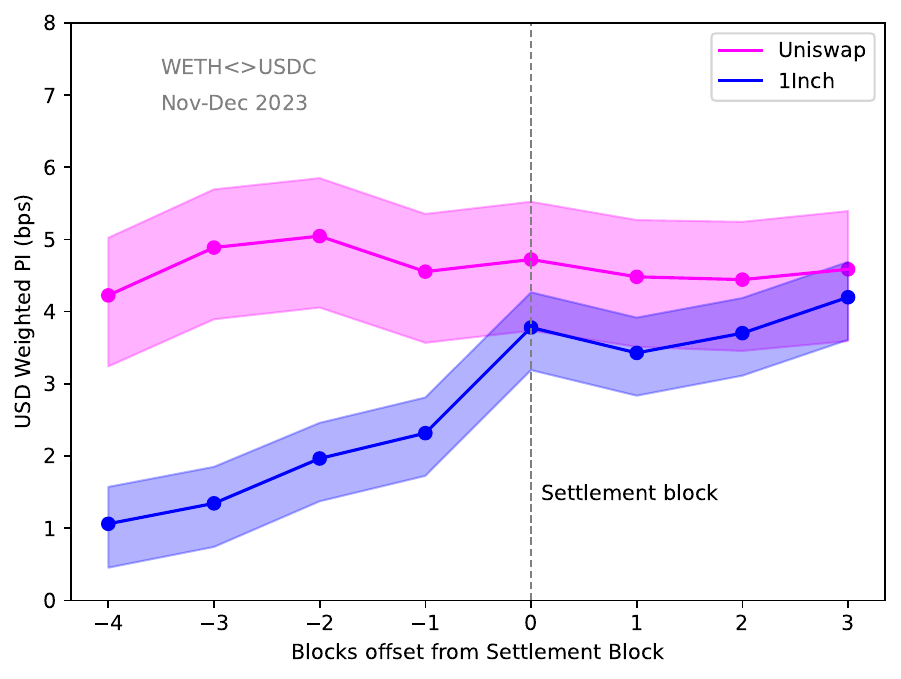}
    \end{subfigure}
    \caption{This figure presents the USD-weighted price improvement trajectory, $\rho$, for various trading platforms across different time blocks relative to the settlement block. It offers a direct comparison of performance with Uniswap Classic, Uniswap X, 1Inch Aggregator, and 1Inch Fusion, illustrating the fluctuations and trends in price improvements over time. The shaded areas indicate the range of statistical and systematic uncertainty, underscoring the variability in the data.}
    \label{fig:mo_curve}
\end{figure}

In \Cref{fig:mo_curve}, we present an analysis of price improvement using the USD-weighted values computed for trades across various platforms. The left plot in \Cref{fig:mo_curve} depicts these weighted price improvements for each trade settlement channel — 1Inch Aggregator, 1Inch Fusion, Uniswap Classic, and UniswapX — at different time offsets from the settlement block, calculated as per \eqref{eq:pi-mo}. For instance, the initial point on the pink line for Uniswap X indicates an average price improvement of slightly over 5 basis points when compared to a hypothetical trade on Uniswap Classic settled at the end of the fourth block prior to the actual settlement block.

One important case to consider is when the offset is zero, where the counterfactual trade is simulated against the end of the settlement block. This scenario likely mirrors the true market price, as it is presumed most arbitrage would be executed by this point. Examining the results at this offset provides a reasonable assessment of fair price improvement (within the limits of arbitrage created by frictions such as trading fees or gas). Interestingly, trades settled through both 1Inch Fusion and UniswapX show significant price improvements, while those through 1Inch Aggregator and Uniswap Classic do not.  The fact that Uniswap Classic, which purely routed to onchain liquidity, achieves performance statistically indistinguishable from zero at the zero offset, speaks to the accuracy of our counterfactual baseline.

Notably, the data for Uniswap Classic and 1Inch Fusion show some sensitivity to the offset value, which merits further investigation for understanding this variability.
Given the  sensitivity and the previous arguments, we pick the offset $0$ to do further analysis on, which is the most optimistic for both 1Inch and Uniswap: at offset $-1$, the aggregate difference in \Cref{fig:mo_curve} is already statistically significant, and less interesting from a comparative and robustness checking sense.

It is important to remember that a user cannot self-select into one of these settlement channels in the left plot of \Cref{fig:mo_curve} and expect to receive execution that are on par with the sample statistics shown in the left plot, because the samples here contains selection bias embedded in the default settlement path recommendation algorithm implemented by the interface.

To address potential selection biases, we present a comparison at the interface level in the right plot of \Cref{fig:mo_curve}. This approach considers all samples generated by the interface collectively, as interfaces do not actively choose their users, unlike their settlement path selection algorithms. Therefore, inspecting the aggregate performance at the interface level eliminates selection biases and reflects the outcome a user would typically expect without altering default settings. As an illustration, at offset zero, users of the Uniswap interface can expect to receive about 4.6 basis points in price improvement, whereas users of the 1Inch interface can expect around 4.3 basis points. Both interfaces have price improvements significantly above zero, and this result remains robust against various offsets from settlement blocks. This shows that by integrating additional liquidity and implementing additional routing mechanisms, both interfaces are able to provide their users with better execution that our chosen baseline.

The shaded areas in both plots represent the uncertainty regions, the methodology of which is detailed in \Cref{sec:uncertainty}. Non-overlapping uncertainty bands between two points indicate a significant (statistically and systematically) difference in their price improvements. In the case of the right plot of \Cref{fig:mo_curve}, it shows that other than in block offset -4 to -1, Uniswap interface has significantly (statistically and systematically) more price improvements than 1Inch interface, and both interfaces have significantly positive price improvement compared to the common baseline across all 8 offsets.

\begin{figure}[ht]
    \centering
    \includegraphics[width=1\textwidth]{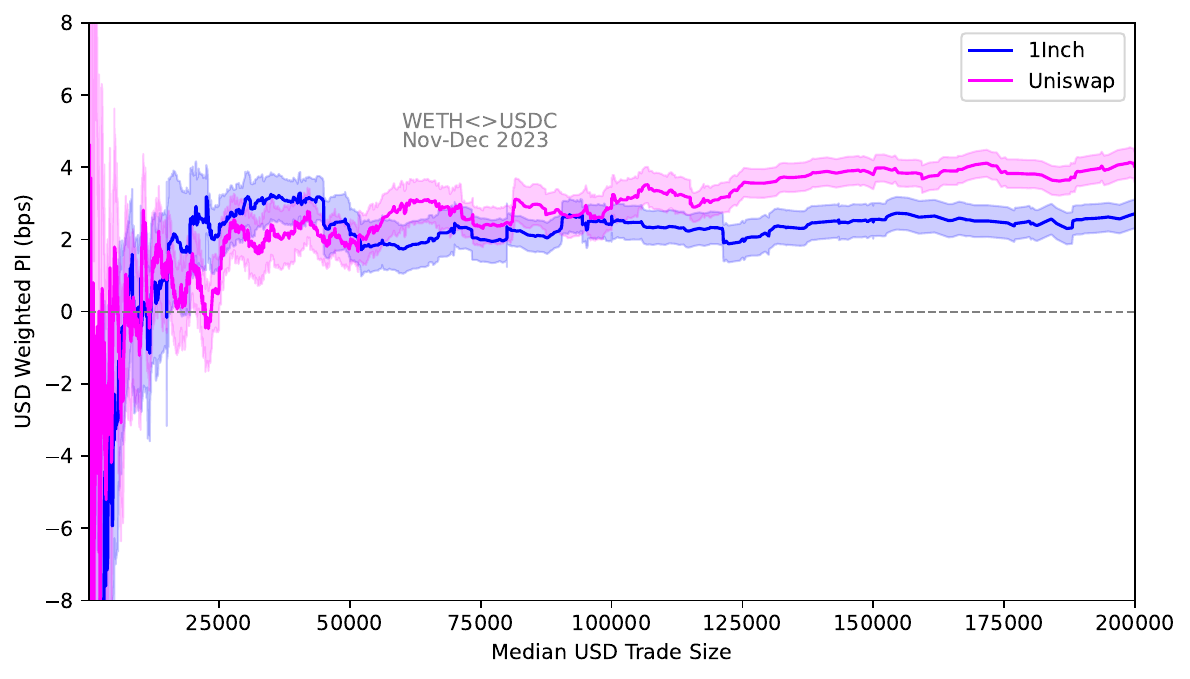}
    \caption{Total Price Improvement (PI) trend analysis for Uniswap and 1Inch platforms, showing the rolling USD-weighted PI across varying median USD trade sizes. The shaded regions represent statistical uncertainty, illustrating the variability and confidence of PI estimates over a wide range of trade sizes.}
    \label{fig:pi_by_company}
\end{figure}

This analysis also allows the examination of how price improvements relate to trade size. To accomplish this, we compute a rolling USD weighted price improvement, across the two interfaces, along with the uncertainties. The results are summarized in \Cref{fig:pi_by_company}. We see that for small trade sizes (i.e. when USD size is < \$25k), the average price improvement is very noisy, but as trade size increases, price improvement stabilizes and is consistently positive. Average price improvement on 1Inch stablizes around 2 basis points, whereas price improvement on Uniswap continues to grow to almost 4 basis point when median trade size is about \$200k.

\begin{figure}[ht]
    \centering
    \includegraphics[width=1\textwidth]{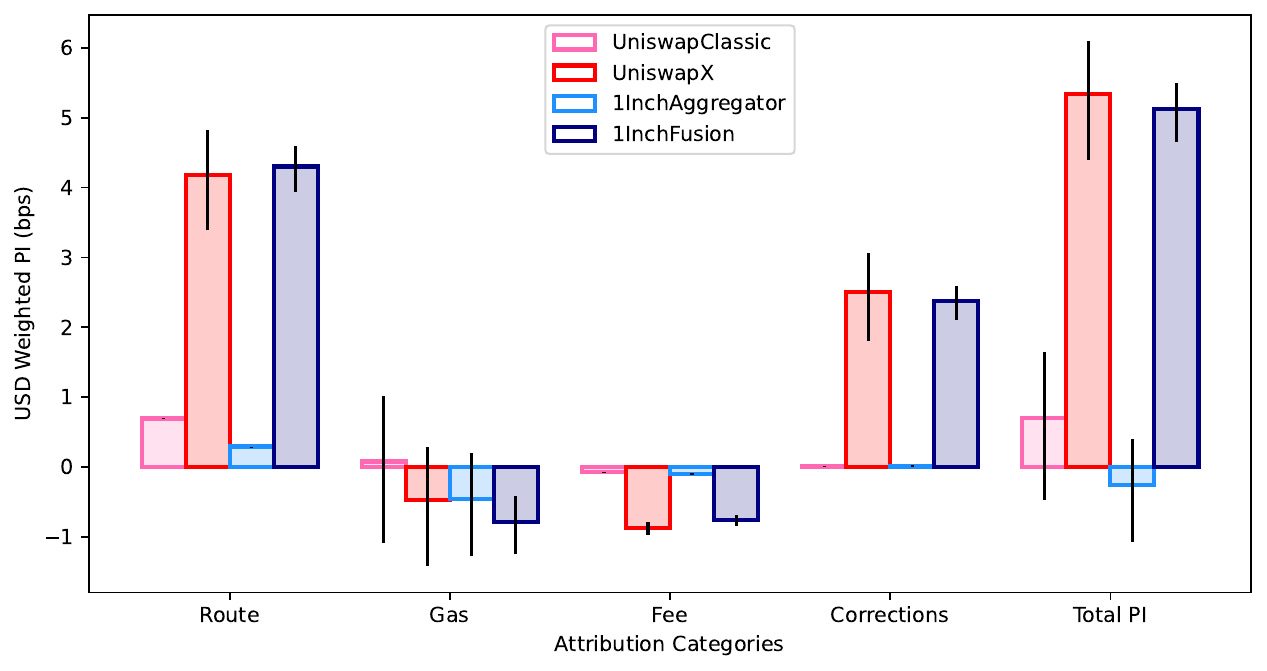}
    \caption{A breakdown of the USD-weighted price improvement (PI) contributions for all four settlement paths. The chart presents a comparative view of routing, gas, and fee contributions to overall PI, along with the necessary corrections. The total PI values are represented, highlighting the differential impact of each component on the respective platforms.}
    \label{fig:attribution_summary}
\end{figure}

Next, as described in \Cref{subsec:attribution}, the method facilitates an attribution of price
improvement to the various components of OFA controllable quantities. For all four settlement
paths, we perform this attribution at settlement time (offset $0$ in the left panel of \Cref{fig:mo_curve}). The results are summarized in \Cref{fig:attribution_summary}.

There are several interesting patterns in the attribution result. First, we can see that the attribution is a good approximation by verifying that the corrections, terms $\mathcal{O}(\Delta^2)$, were indeed small for 1Inch Aggregator and Uniswap Classic. Second, the majority of price improvements on both interfaces is achieved through better routing in their respective OFAs, or in other words, added access to liquidity sources and providers. Third, both OFAs (1Inch Fusion and UniswapX) as well as 1Inch Aggregator show small but non-zero amounts of gas overhead compared to Uniswap Classic, which translate to roughly 0.5-1 basis points of degradation in execution quality. Fourth, both OFAs have a decent amount of price improvements achieved through the correction term, which economically can be interpreted as the interaction term between gas efficiency and priority fee savings. In general, the 1Inch and the Uniswap interfaces demonstrate similar underlying economic dynamics, and price improvements are shown to be attributable to similar mechanisms. Both are significantly positive.

\begin{figure}[ht]

    \centering
\includegraphics[trim=0 100 0 150, clip, width=1\textwidth]{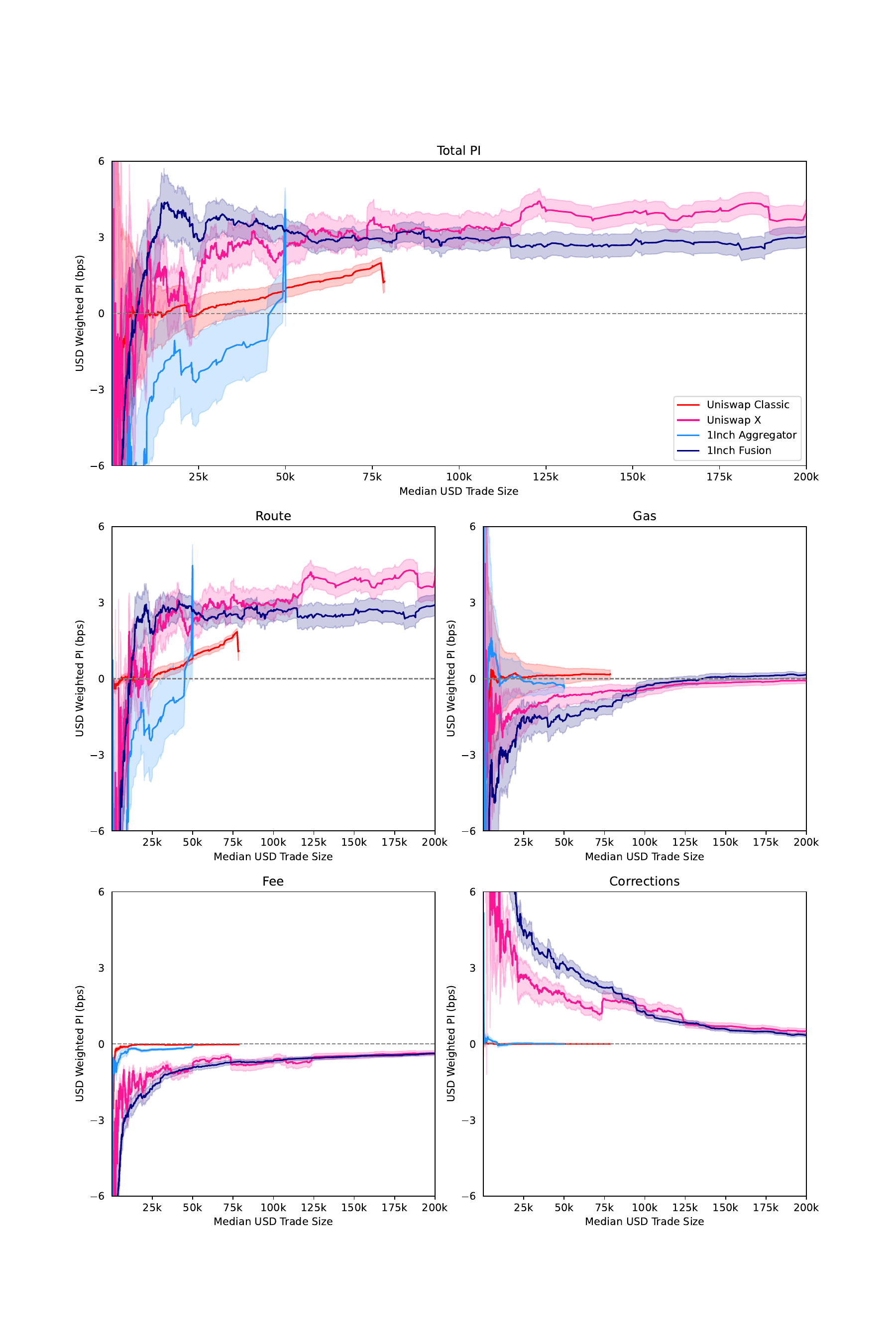}

    \caption{Comparative analysis of price improvement (PI) contributions by trade size for Uniswap Classic, 1Inch Aggregator, Uniswap-X and 1Inch Fusion platforms. The graphs display the overall PI, corrections to PI, and individual contributions from gas optimization and routing efficiency, as well as the impact of priority fee settings. Trends indicate how each component influences the price improvement across varying median USD trade sizes, offering insights into the optimization strategies and cost-efficiency of each trading interface}
    \label{fig:attribution_by_usd}

\end{figure}

Lastly, to highlight the additional insights of the attribution model, we further attribute PI across different median trade sizes groups into modifiable components, as shown in \Cref{fig:attribution_by_usd}. First, note that the corrections $\approx 0$ for Uniswap Classic and 1Inch Aggreagator, giving validity to our linear approximations in certain regions. Next, we can see as trade size increases, both 1Inch Aggregator and Uniswap Classic provide better output performance, with 1Inch's increase surpassing Uniswap's for very large USD. However, we also see that 1Inch's Aggregator gas efficiency decreases for large orders, whereas, Uniswap Classic's is consistent across all trade sizes.

The corrections for Uniswap-X and 1Inch Fusion are higher. This indicates that the non-linear terms in the Taylor series are contributing significantly to the PI. The attribution shows that 3-4 bps of PI consistently comes from Routing, which in this case, is interpreted as the ability to access more liquidity. The other non-linear terms are interactions between gas and priority fee. We suspect that the mechanisms by which these are internalized are more complicated, leading to non-linear effects. We hope to understand this more in the future.


\section{Outlook}

This research has demonstrated a proof of concept for our framework designed to analyze and optimize onchain order flow auctions. Our approach focuses on price improvement and provides a model to attribute performance gains to specific components such as routing efficiency, gas optimization, and priority fee adjustments.

Looking ahead, there are several directions for further development and application of our work. Our methodology is particularly suitable for evaluating OFAs where gas costs are internalized, like those used by platforms such as 1Inch Fusion and UniswapX. By refining our attribution model (looking at non-linear corrections), we aim to provide deeper insights into how these systems manage transaction costs and improve trading outcomes, potentially influencing decentralized finance standards. In addition, we hope to further look at how on-chain versus off-chain fillers perform in this attribution, giving insights into future RFQ-like design choices. 

Furthermore, our framework's flexibility make it applicable to other trading mechanisms, including batched auctions like COW Swap and rebate systems like MEV Share. For batched transactions, we plan to internalize the gas cost of the batch into the realized prices, thereby, understanding how much gas savings and uniform clearing prices impact a single transaction's PI. For rebate-like mechanisms, we hope to understand how much PI can be attained from sources like back-running. Extending the empirical analysis to a larger set of OFA systems would be helpful for the community in terms of guiding users to the most efficient platform.

In summary, our research enhances understanding of blockchain-based trading systems and establishes a foundation for future studies that could improve the efficiency, transparency, and fairness of decentralized financial markets. The ongoing refinement and application of this methodology will contribute to the development and sophistication of cryptocurrency trading strategies, benefiting various stakeholders in the blockchain ecosystem.

\clearpage

\appendix

\section{Price Details}\label{sec:price_details}

The computation of realized and counterfactual prices change depending on if gas is/not internalized and also whether the input toke is/not WETH. In this section, we describe in detail how the prices are calculated.

\paraheader{Realized prices.}
When gas is not internalized (Uniswap Classic, 1Inch Aggregator), we have two cases,
\begin{equation}
        p = \begin{cases}
                \frac{o - g(b+f)}{i} \,, \quad \text{when token out address = \WETH}\\
                \frac{o}{i+g(b+f)} \,, \quad\text{when token in address = \WETH} \\
        \end{cases} \,.
\end{equation}

When gas is internalized (UniswapX, 1Inch Fusion), we compute the realized price,
\begin{equation}
        p = \frac{o}{i} \,.
\end{equation}

\paraheader{Counterfactual prices.}
As mentioned before, the baseline function generates quotes
\begin{equation}
        \mathcal{B}(i) \rightarrow (o^\prime, g^\prime) \, .
\end{equation}
When the API is given an input $i$, it can generate a token out amount estimate $o^\prime$, and gas use estimate $g^\prime$.

When the gas is not internalized (Uniswap Classic, 1Inch aggregator), we compute $p^\prime$ as
\begin{equation} \label{eq:cf_price_gasNotInternalized}
        p^\prime = \begin{cases}
                \frac{o^\prime - g^\prime(b + f^\prime)}{i} \,,\quad \text{when token out address = \WETH}\\
                \frac{o^\prime}{i+g^\prime (b + f^\prime)} \,, \quad \text{when token in address = \WETH} \\
        \end{cases} \,.
\end{equation}

When gas is internalized (UniswapX, 1Inch Fusion), we compute a token in amount gas adjusted $i^\prime = i - g^\prime ( b + f^\prime)$,
we define the following
\begin{equation}
        \mathcal{B}(i^\prime) = (o^{\prime\prime}, g^{\prime\prime})\,,
\end{equation}
and
\begin{equation} \label{eq:cf_price_gasInternalized}
        p^\prime = \begin{cases}
                \frac{o^\prime - g^\prime(b + f^\prime)}{i} \,, \quad \text{when token out address = \WETH}\\
                \frac{o^{\prime\prime}}{i^\prime + g^\prime (b + f^\prime)} \,, \quad \text{when token in address = \WETH} \\
        \end{cases} \,.
\end{equation}
Note that the denominator is just $i$, but we have written it in a way that allows us to estimate the amount of input token that you have available to route $i^\prime$, and the amount that you must pay in gas $g^\prime (b + f^\prime)$. The amount of token that you have available to route is then used to generate the token out amount estimate $o^{\prime\prime}$.


\section{Simulation Process}\label{sec:simulation_process}

The process for generating our baseline for each swap is outlined in detail below:

\begin{enumerate}
    \item \textbf{\sffamily API Call for Route Calculation}: For each completed swap, we initiate a request to the routing API, instructing it to find the best possible route for a swap with an exact input amount matching that of the realized swap. To ensure a comprehensive analysis, the routing API is supplied with the state of the blockchain for several blocks surrounding the settlement block---specifically, $n$ blocks before and after. This approach allows us to perform a robustness check by considering potential variations in the blockchain state at times slightly offset from the actual transaction block.
    
    \item \textbf{\sffamily Route Identification and Calldata Formation}: For each specified block, the routing API assesses the state of the blockchain as of the end of that block. It also takes as input some statistics related to liquidity pools, such as TVL (total value locked). These information is retrieved from subgraph on The Graph and supplied to the routing API. The API then determines the optimal route through all available Uniswap pools that maximizes the output token amount, net of gas fees. The selected route is then formatted into calldata, which are the instructions necessary for the blockchain's execution client to perform the swap. Additionally, the API provides an estimate of the gas costs associated with executing the transaction.
    
    \item \textbf{\sffamily Simulation via Tenderly}: The calldata obtained from the routing API are then input into Tenderly \cite{Tenderly2024}, a blockchain execution simulator. To maintain consistency in our simulations, we set a priority fee of 0.1 Gwei. We also pass in the block number for the simulation. Tenderly processes these inputs and returns the simulated output, detailing the amount of gas consumed and providing additional metadata related to the simulation.
\end{enumerate}

This detailed simulation process ensures that our baseline is not only reflective of the optimal outcomes possible under varying blockchain states but also robust and reproducible, providing a solid foundation for comparing the efficacy of different Order Flow Auction implementations.

{\small
  \singlespacing
  \bibliographystyle{unsrt}
  \bibliography{references}
}

\end{document}